\pdfoutput=1

\documentclass[11pt]{article}

\usepackage{acl}

\usepackage{times}
\usepackage{latexsym}
\usepackage{url}            %
\usepackage{booktabs}       %
\usepackage{amsfonts}       %
\usepackage{nicefrac}       %
\usepackage{microtype}      %
\usepackage{xcolor}         %
\usepackage{graphicx}
\usepackage{multirow}
\usepackage{multicol}
\usepackage{subcaption}
\usepackage[most]{tcolorbox}

\usepackage[T1]{fontenc}

\usepackage[utf8]{inputenc}

\usepackage{microtype}

\usepackage{inconsolata}

\newcommand{\datasetname}{NevIR}

\title{\datasetname{}: Negation in Neural Information Retrieval}

\author{Orion Weller, Dawn Lawrie, Benjamin Van Durme \\
        Johns Hopkins University \\   \texttt{oweller@cs.jhu.edu}}

\begin{document}
\maketitle
\begin{abstract}
Negation is a common everyday phenomena and has been a consistent area of weakness for language models (LMs).
Although the Information Retrieval (IR) community has adopted LMs as the backbone of modern IR architectures, there has been little to no research in understanding how negation impacts neural IR.
We therefore construct a straightforward benchmark on this theme: asking IR models to rank two documents that differ only by negation.
We show that the results vary widely according to the type of IR architecture: cross-encoders perform best, followed by late-interaction models, and in last place are bi-encoder and sparse neural architectures.
We find that most information retrieval models (including SOTA ones) do not consider negation, performing the same or worse than a random ranking.
We show that although the obvious approach of continued fine-tuning on a dataset of contrastive documents containing negations increases performance (as does model size), there is still a large gap between machine and human performance.\footnote{Code and data are available at \url{https://github.com/orionw/NevIR}}
\end{abstract}

\section{Introduction}
Recent work in natural language processing (NLP) has shown that language models (LMs) struggle to understand text containing negations \cite{ravichander2022condaqa,mckenzie2022round1} and have poor performance compared to humans. This unresolved problem has downstream implications for information retrieval (IR) models, which use LMs as the starting backbone of their architectures. 

\begin{figure}[t]
\centering
    \includegraphics[trim={0cm 3.75cm 0cm 0cm},clip,width=\linewidth]{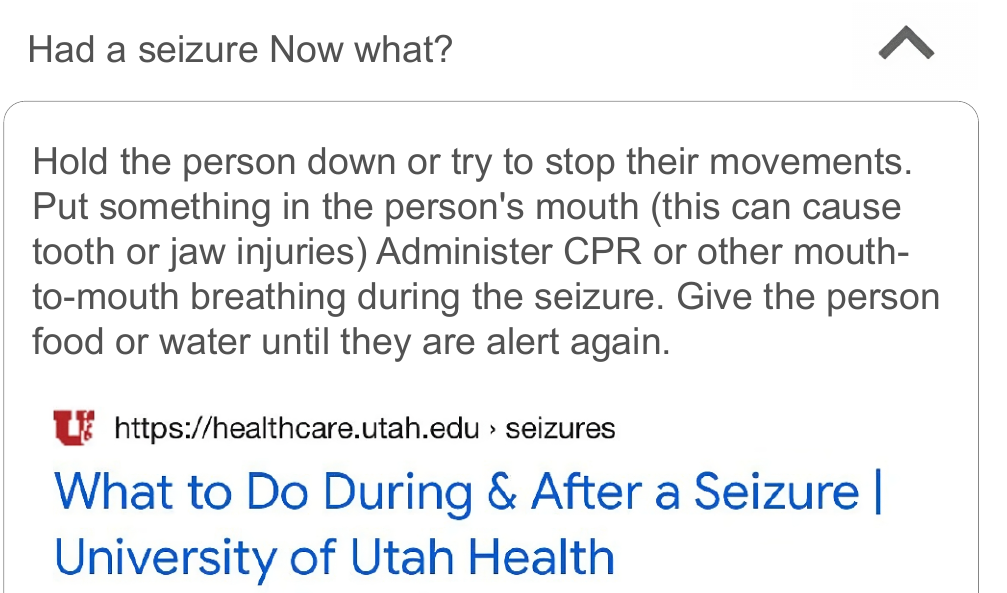} 
    \caption{Negation is something not well understood by IR systems. This screenshot shows Google Search making a deadly recommendation because of its failure to catch the negation in the article (e.g. ``do not ...").\label{fig:tweet}\vspace{-1.25em}}
\end{figure}

However, work on negation in IR has mainly focused on pre-neural (e.g. no LM) retrieval \cite{whan1990model,mcquire1998ambiguity,averbuch2004context,kim2019statute}, with no research into how negation affects modern neural IR. This failure to understand negation in IR can lead to devastating consequences in high stakes situations, like the prominent case where Google Search told users what to do during a seizure by listing off bullet points from a website that was specifically specifying what \textbf{not} to do (Figure~\ref{fig:tweet}). One can easily imagine other serious failure cases in high-stakes domains such as law, education, or politics. Even for casual everyday usage, a lack of understanding of negation by neural IR ignores an entire category of user queries, such as ``Where should I not stay in [vacation town]?", ``Who did not win an Oscar in 2023?", or ``What information has OpenAI failed to release about GPT-4?"

\begin{figure*}[t]
    \centering
    \includegraphics[trim={0cm 0cm 0 0cm},width=\linewidth]{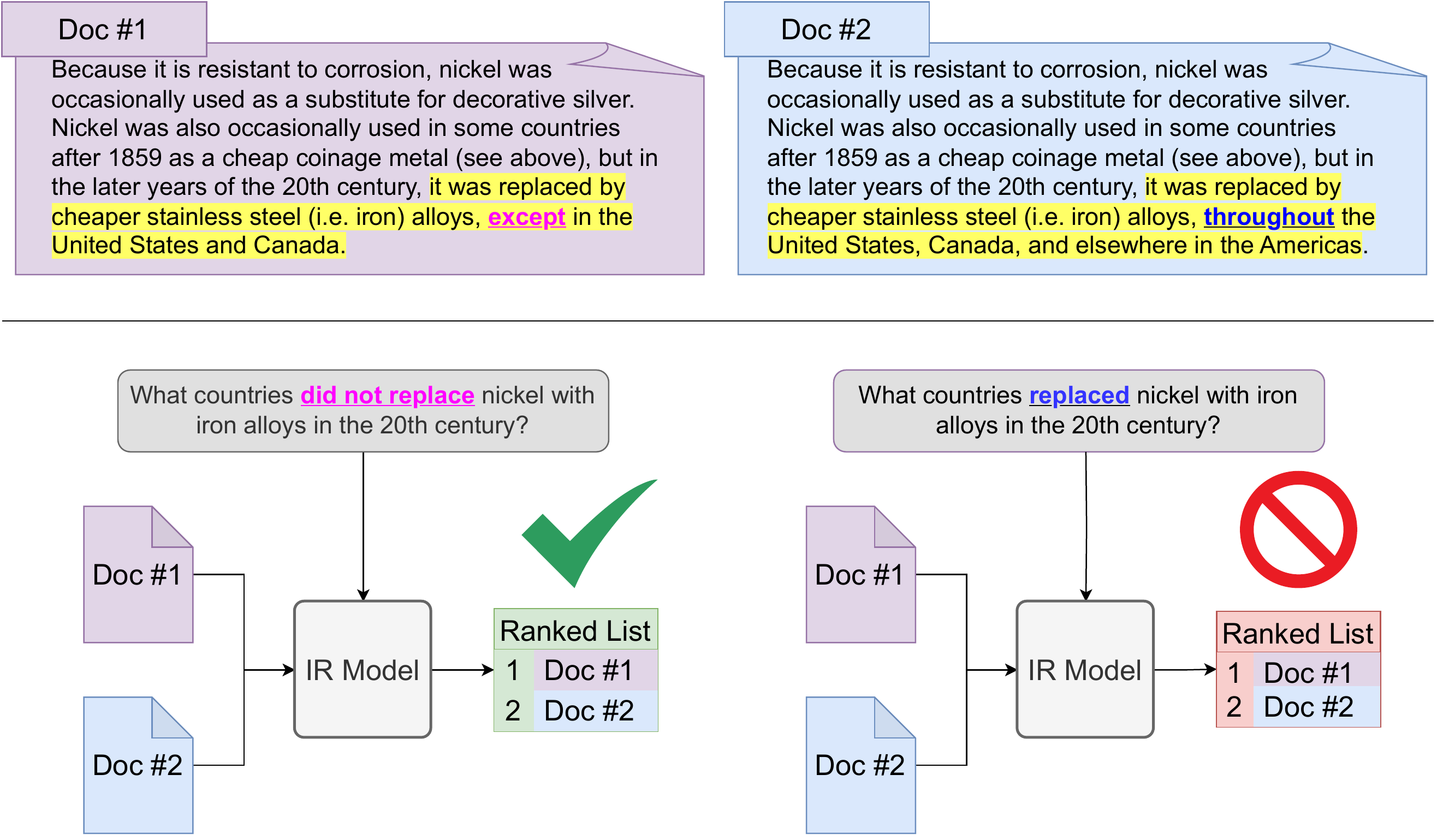}
    \caption{An example instance and the evaluation process. The initial documents from CondaQA \cite{ravichander2022condaqa} are used to create the queries via Mechanical Turk. The lower half shows the pairwise accuracy evaluation process, where the model must rank both queries correctly. In this example, the IR model scored zero paired accuracy, ranking Doc \#1 above Doc \#2 in both queries (and failing to take into account the negation).\label{fig:example}\vspace{-1em}}
\end{figure*}

We aim to fill this gap in the literature by providing a benchmark for Negation EValuation in Information Retrieval, dubbed \datasetname{} (pronounced ``\textit{never}"). \datasetname{} builds off of existing work in negation \cite{ravichander2022condaqa} by using 2,556 instances of contrastive document pairs that differ only with respect to a crucial negation. We then crowdsource query annotations for the two documents in each pair, where each query is only relevant to one of the respective documents and is irrelevant to the other document (Figure~\ref{fig:example}). By doing so, we can test whether models correctly rank the documents when accounting for the negation.

We find that nearly all IR systems ignore the negation, generally scoring one document of the two higher for both queries. Furthermore, state-of-the-art models perform nearly the same or much worse than randomly ranking the document pairs. We provide analysis of these results, showing that bi-encoder representations of the two documents are nearly identical despite negation words and that late-interaction models such as ColBERT ignore negation words in the MaxSim operator.

We also show that continued fine-tuning of IR models on negation data provides some gains on \datasetname{}, but still leaves significant room to improve (while also slightly hurting performance on traditional benchmarks such as MSMarco). We hope that our analysis will spur increased attention to the problem of negation in information retrieval and provide a dataset for IR training and evaluation.

\section{Background}
\subsection{Motivation}
Information Retrieval (IR) is a broadly defined task of finding relevant pieces of information based on a query in natural language. The specifics of IR can vary broadly across languages, domains (e.g. legal), and purposes (e.g. counterarguments, lists, general factoids). Note that many of these specialized cases would be improved through a better understanding of negation, such as lists, counterarguments, and domain-specific language (e.g. legal or medical).

Along with the improvement from neural IR, there has been a surge of interest in retrieval-augmented language models, such as RAG \cite{lewis2020retrieval}, FiD \cite{izacard2020leveraging}, and SeeKeR \cite{shuster2022language}. In just the last few months, generative retrieval has been productionized, with systems such as Google's Bard, Bing Chat, and You.com.\footnote{\href{https://bard.google.com/}{https://bard.google.com/}, \href{https://www.bing.com/new}{https://www.bing.com/new}, and \href{https://you.com}{https://you.com}} These systems combine IR models with large language models, enabling them to find and generate responses to queries on the fly. 

Thus, as LMs and IR systems become more intertwined and used in production, understanding and improving their failure cases (such as negation) becomes crucial for both companies and users. 

\subsection{Neural IR}
Since 2020, neural models for information retrieval have generally outperformed traditional sparse methods (such as BM25) in most situations \cite{karpukhin-etal-2020-dense,khattab2020colbert}. Given a large collection of training data, these models are optimized using a contrastive loss in order to learn how documents are related to a given query. These methods provide several advantages over sparse methods, including the ability to go beyond simple lexical matches to encode the semantic similarity of the natural language text. 

Recent work has focused on the ability of neural models to generalize to new domains, without any domain-specific training data (e.g. zero-shot). One prominent benchmark for this type of work is the BEIR dataset suite \cite{thakur2021beir} which evaluates models' generalization on a range of diverse IR datasets. Our work provides both zero-shot (no model fine-tuning) and standard train/test splits to accommodate both paradigms.

\subsection{Negation in NLP}
Negation has also been an area where LMs typically perform below average \cite{li-huang-2009-sentiment,he-etal-2017-neural,hartmann-etal-2021-multilingual,ettinger2020bert}. Recent work on negation in NLP has shown that although LMs struggle with negation, it does improve with model scaling and improved prompting techniques \cite{mckenzie2022round1,wei2022inverse}. Despite scale improvements, these works (and other follow up works, c.f. \citet{ravichander2022condaqa,hossain-etal-2022-analysis}) have shown that LMs still struggle with negation and are in need of new datasets and methods to improve performance.

As modern IR models use LMs as the backbone of their architectures, it is intuitive that negation will pose problems to IR systems as well. This problem is compounded as IR models are not able to scale to larger LMs as easily, due to efficiency and latency constraints on processing large amounts of documents in real-time.

\subsection{Negation in IR}
Negation has been a weak point for information retrieval methods throughout the years. Early work in information retrieval \cite{whan1990model,strzalkowski1995natural} has demonstrated the difficultly of negation for non-neural methods like TF-IDF \cite{sparck1972statistical} and BM25 \cite{robertson1995okapi} when used out of the box.

To the best of our knowledge, there is little to no published work on negation for neural models. The most similar area in IR is that of argument retrieval \cite{wachsmuth-etal-2018-retrieval,bondarenko2022overview}, also included in the BEIR dataset, whose aim is to find a counterargument for the given query. However, these datasets implicitly ask the model to find the counterargument to the query through the task design and specifically don't include negation in the query. So although argument retrieval datasets contain a larger amount of negations compared to standard IR datasets like MSMarco \cite{nguyen2016ms}, negation is not a conscious choice in the design of either the documents or the queries and is confounded by the implicit task definition. In contrast, we explicitly provide and measure the impact of negation on both documents and queries. 

Another recent work by \citet{opitz2022sbert} incorporates features from Abstract Meaning Representation (AMR) parsing (including negation, as one of many) to improve SBERT training. However, they only evaluate negation for AMR parsing (and on AMR datasets) whereas we focus on negation in IR and create a benchmark for ranking.

\subsection{Contrastive Evaluation}
\label{sec:contrastive}
Contrastive evaluation has emerged as a promising evaluation technique: constructing datasets that consist of minor differences but that test crucial distinctions \cite{gardner2020evaluating,kaushik2019learning}. For IR specifically, this has included testing sentence order \cite{rau2022role}, lexical structures \cite{nikolaev2023representation},  general axiom creation \cite{volske2021towards}, paraphrases, mispellings, and ordering \cite{penha2022evaluating}, LLM-based query and document expansion \cite{weller2023generative}, and factuality, formality, fluency, etc. \cite{macavaney2022abnirml}. We follow these works by evaluating not on a classical IR evaluation corpus, but rather with paired queries and documents.

\section{Creating \datasetname{}} 
We test negation in neural IR using a contrastive evaluation framework, which has shown great utility in understanding neural models (Section~\ref{sec:contrastive}).

\subsection{Contrastive Documents}
\begin{figure*}[t]
    \centering
    \includegraphics[trim={0.5cm 1cm 0 0cm},width=0.5\linewidth]{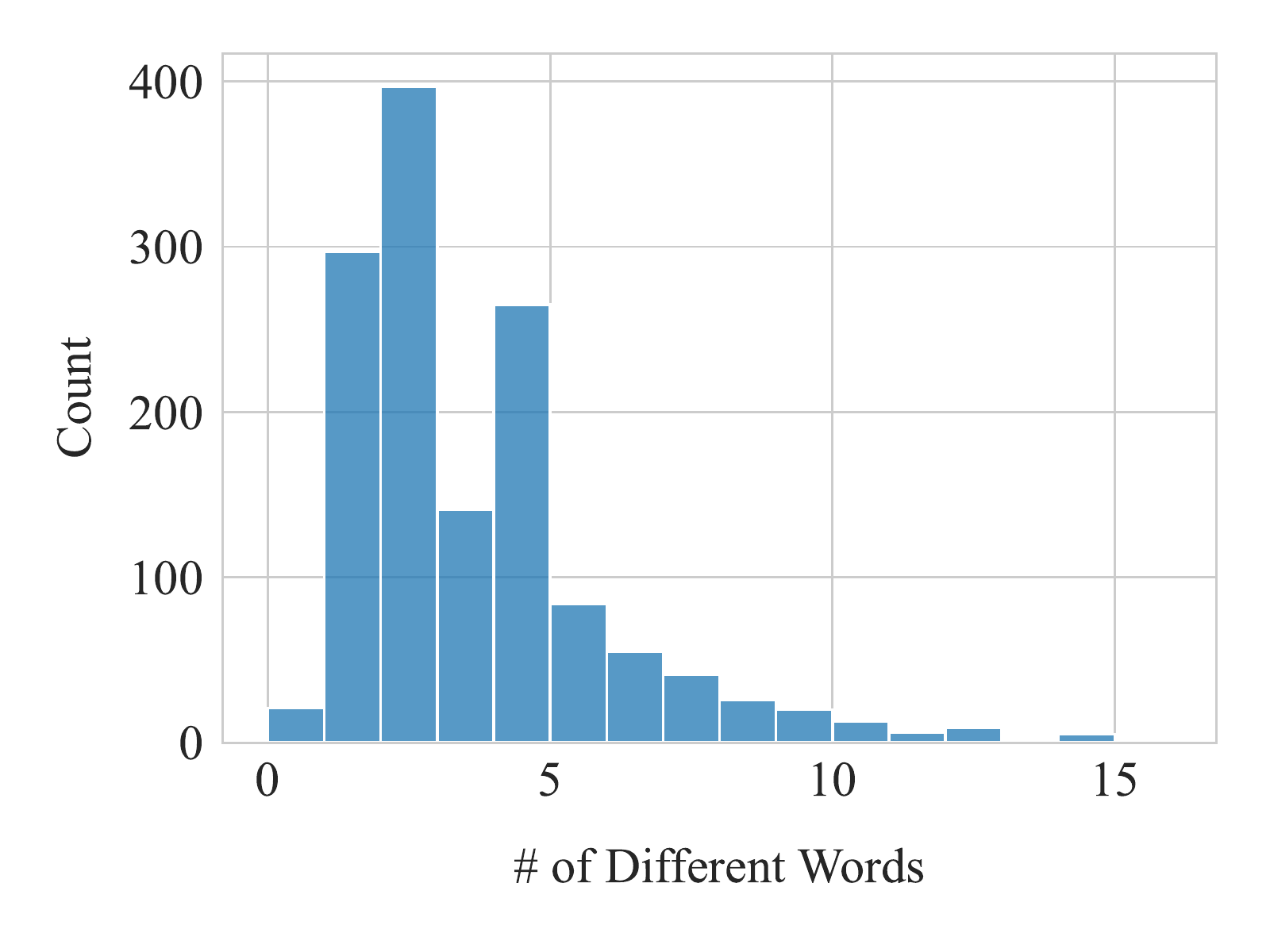}
    \includegraphics[trim={0.5cm 1cm 0 0cm},width=0.49\linewidth]{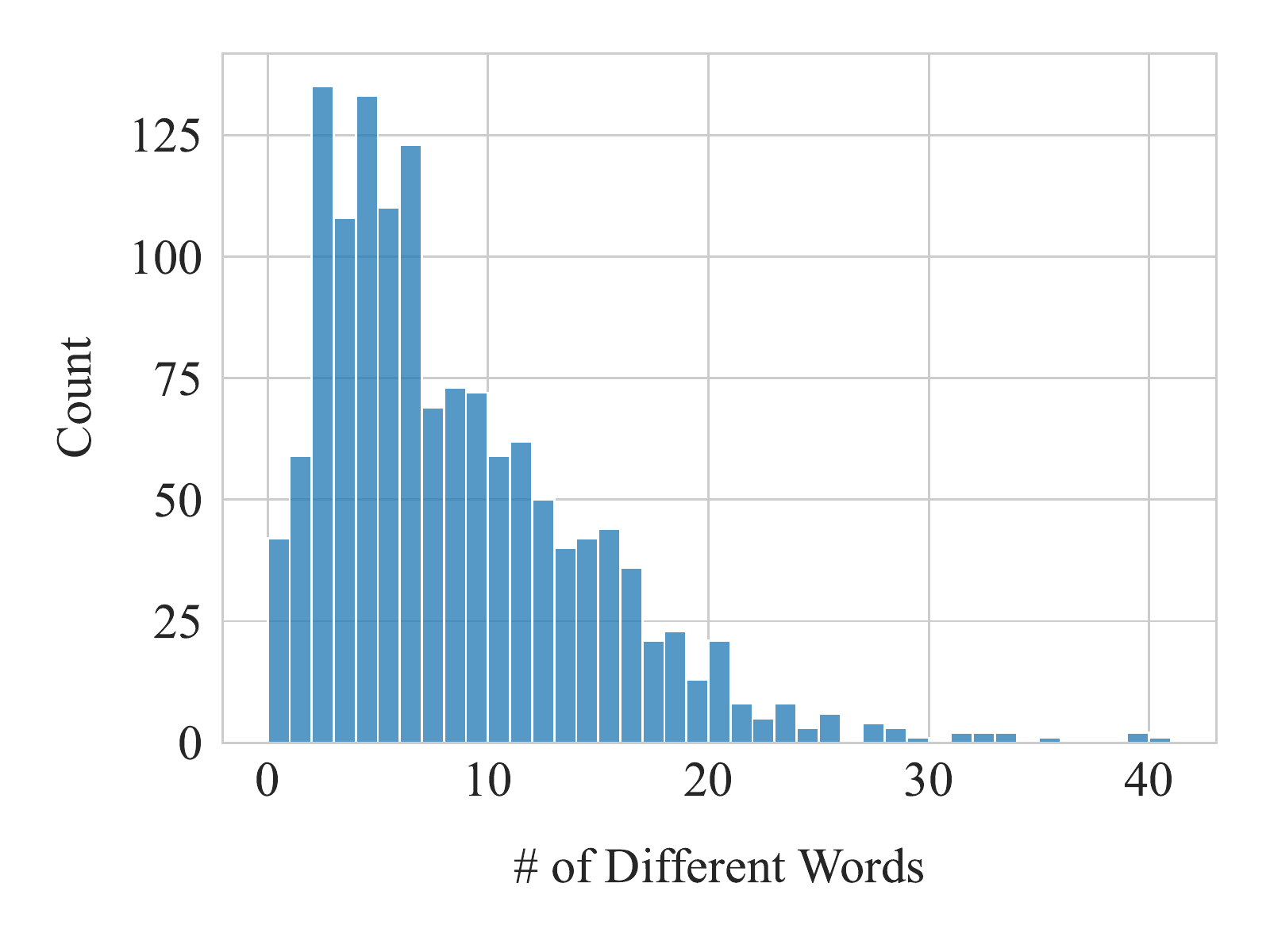}
    \caption{The distribution of the number of different (e.g. unique) words between the queries (left) or documents (right) in each pair. The average length differences are shown in Table~\ref{tab:statistics}.\label{fig:diff_words}\vspace{-1em}}
\end{figure*}

We start by collecting pairs of documents that differ as minimally as possible but include negation, using the CondaQA \cite{ravichander2022condaqa} dataset as a starting point. CondaQA consists of ``in-the-wild" natural paragraphs that contain negation and human-edited versions of those paragraphs that either paraphrase, change the scope of the negation, or undo the negation. For our IR benchmark, we exclude the paraphrase edits, as they do not provide different semantic meanings for comparison. Thus, this allows us to compare the effect of the negation between document pairs with a minimal lexical difference (see Table~\ref{tab:statistics} and Figure~\ref{fig:diff_words} for statistics).

\begin{table}[t]
\centering
\begin{tabular}{lrrr}
\toprule 
{\bf Statistic} & {\bf Train} & {\bf Dev} & {\bf Test} \\
\midrule 
\# Pairs & 948 & 225 & 1383 \\
\midrule
Question 1 Length & 10.9 & 11.1 & 11.0 \\
Question 2 Length & 11.2 & 11.4 & 11.4 \\
Average Length Diff & 0.95 & 1.05 & 1.01 \\
\midrule
Document 1 Length & 112.5 & 113.0 & 113.7 \\
Document 2 Length & 115.6 & 116.8 & 116.8 \\
Average Length Diff & 4.39 & 4.71 & 4.16 \\
\bottomrule
\end{tabular}
\caption{\datasetname{}\ statistics, where length is measured in words. Note that the average length differences only take into account total length; for the distribution of unique word differences see Figure~\ref{fig:diff_words}.\label{tab:statistics}\vspace{-0.75em}}
\end{table}

\subsection{Collecting Contrastive Queries}
\label{sec:design}
To test whether IR models correctly rank the documents, we collect natural language queries for those document using workers on Amazon's Mechanical Turk. We ask workers to create one query for each of the two paragraphs, with four constraints:
\begin{enumerate}
    \setlength\itemsep{0em}
    \item The answer to the queries are the same for both paragraphs
    \item The question is answered by a span (e.g. not a yes/no or boolean answer)
    \item The question contains enough information to identify the relevant passage from a collection of documents (e.g. it contains relevant entity names, not just ``when was he born?")
    \item The question can \textbf{only be answered by one} of the two paragraphs (thus making the other paragraph irrelevant)
\end{enumerate}

Note that boolean questions would be relevant to both documents, and hence they were excluded. To help annotators understand the task, we allowed them to test their queries against a small neural cross-encoder model (\textit{all-mpnet-base-v2} from \citet{reimers-2019-sentence-bert}) but did not require them to. The annotation interface is in Appendix~\ref{app:annotation_interface}. 

Through a series of initial pilot HITs, we found that annotators would typically quote verbatim from the passage and use the words that were only present in only one document. To prevent models from exploiting this shallow heuristic, we included a 5th constraint: not allowing workers to use any word in the query that was only present in one of the two documents. Note that this was an effective but not perfect constraint (as is shown by TF-IDF's 2\% performance in Table~\ref{tab:main}), as any non-exact string match including subwords, plural versions, etc. would pass this validation check.

We recruited annotators with greater than 99\% HIT acceptance rate and greater than 5000 completed HITs.  All annotators participated in two paid trial HITs where their work was assessed before moving on. Workers were paid \$2.5 USD for approximately six minutes per HIT, for an average of \$15 USD per hour. Overall, we had 28 unique annotators with an average of 91 query pairs each.

\begin{table*}[t]
\centering
\begin{tabular}{p{2.5cm}p{1.5cm}p{1cm}p{7.5cm}p{1cm}}
\toprule 
 Type & Data & Params & Model Name & Score \\
\midrule 
Random & N/A & 0 & Random & 25\% \\
\midrule 
 \multirow{3}{*}{Sparse} 
& N/A & N/A & TF-IDF \cite{scikit-learn} & 2.0\% \\
& MSMarco & 110M & SPLADEv2 ensemble-distill \cite{10.1145/3477495.3531857} & 8.0\% \\
& MSMarco & 110M & SPLADEv2 self-distill \cite{10.1145/3477495.3531857} & 8.7\% \\
\midrule
 \multirow{2}{*}{Late Interaction} 
& MSMarco & 110M & ColBERTv2 \cite{santhanam2022colbertv2} & 13.0\% \\
& MSMarco & 110M & ColBERTv1 \cite{khattab2020colbert} & 19.7\% \\
\midrule
 \multirow{6}{*}{Bi-Encoders} 
& NQ & 219M & DPR \cite{karpukhin-etal-2020-dense} & 6.8\% \\
& MSMarco & 110M & msmarco-bert-base-dot-v5 & 6.9\% \\
& MSMarco & 110M & coCondenser \cite{gao2021unsupervised} & 7.7\% \\
& MSMarco & 85M & RocketQA v2 \cite{ren2021rocketqav2} & 7.8\% \\
& NQ & 66M & nq-distilbert-base-v1 & 8.0\% \\
& MSMarco & 110M & all-mpnet-base-v2 & 8.1\% \\
& MSMarco & 66M & msmarco-distilbert-cos-v5 & 8.7\% \\
& MSMarco  & 170M & RocketQA v1 \cite{qu2020rocketqa} & 9.1\% \\
& QA Data & 110M & multi-qa-mpnet-base-dot-v1 & 11.1\% \\
\midrule
 \multirow{5}{*}{Cross-Encoders} 
& MSMarco & 85M & RocketQA v2 \cite{ren2021rocketqav2} & 22.4\% \\
& STSB & 355M & stsb-roberta-large & 24.9\% \\
& MSMarco & 303M & RocketQA v1 \cite{qu2020rocketqa} & 26.3\% \\
& MSMarco & 61M & MonoT5 small \cite{nogueira-etal-2020-document} & 27.7\% \\
& MNLI & 184M & nli-deberta-v3-base & 30.2\% \\
& QNLI & 110M & qnli-electra-base & 34.1\% \\
& MSMarco & 223M & MonoT5 base (default) \cite{nogueira-etal-2020-document} & 34.9\% \\
& MSMarco & 737M & MonoT5 large \cite{nogueira-etal-2020-document} & 45.8\% \\
& MSMarco & 2.85B & MonoT5 3B \cite{nogueira-etal-2020-document} & 50.6\% \\

\bottomrule
\end{tabular}
\caption{Results for pairwise contrastive evaluation using paired accuracy. All models are from sentence-transformers \cite{reimers-2019-sentence-bert} unless otherwise cited. Data indicates the main source of training data for the model, while score indicates Pairwise Accuracy (see Sec~\ref{sec:evaluation}). Note that RocketQA includes both a cross-encoder and bi-encoder for both versions. TF-IDF scores were designed to be low in the task instruction (Section~\ref{sec:design}).\label{tab:main}\vspace{-1em}}
\end{table*}

\subsection{Dataset Statistics}
Dataset statistics are in Table~\ref{tab:statistics}, showing that the average number of words is around 11 for questions and 113 for documents. The average difference in word length between questions and documents is 1 and 4 respectively, showing that items in each pair are nearly the same length. The distribution of unique word differences between queries and documents is in Figure~\ref{fig:diff_words} and shows that most queries have small differences of 2 to 5 words, although some differ only by a single negation word and some differ by more than five. The difference between the two documents is much more variable, with about 5-10 different words between them.

\subsection{Human Performance}
To verify that this dataset is trivial for humans, we asked three annotators to perform the ranking task on 10 randomly sampled test instances. In all three cases, all human annotators ranked all queries correctly, indicating the simplicity of the task.

\section{Experimental Settings}
\subsection{Metric}
\label{sec:evaluation}
In early investigations we observed that IR models tended to rank one document above the other for both queries. This motivates our usage of a \textit{pairwise accuracy} score to avoid score inflation when models don't actually understand the negation. We start by having the IR model rank both documents for each query. Then, if the model has correctly ranked the documents for both queries (flipping the order of the ranking when given the negated query) we know that the model has correctly understood the negation and the pair is marked as correct. 

\subsection{Models}
We evaluate a wide variety of models in order to show a comprehensive evaluation across common neural IR model types.  We note that although there are other models we do not use (as well as many different strategies for model training), all the major types of retrieval models are accounted for here. We evaluate on the following IR model categories:

\paragraph{Sparse} We evaluate sparse IR models that use the bag-of-words representation during retrieval. This includes TF-IDF (the only non-neural IR method, here as a baseline), and two variants of SPLADE v2++ \cite{10.1145/3477495.3531857,https://doi.org/10.48550/arxiv.2109.10086,10.1145/3477495.3531833}, the ensemble distillation and self-distillation methods. Note that other variants of SPLADE perform worse than these two methods. We do not include BM25 as implementations of BM25 perform similar to TF-IDF due to the small collection and lexical similarity within the pair.

\paragraph{Late Interaction} Late interaction models like ColBERT \cite{khattab2020colbert,santhanam2022colbertv2} embed documents and queries into one vector for each sub-word token. At inference time, these models need to compute a MaxSim operation between query vectors and document vectors to determine similarity. We use both ColBERT v1 and v2 in our experiments.\footnote{We reproduce ColBERT v1 weights from their repository. We do not use PLAID \cite{santhanam2022plaid} or quantization as there are only two documents in the collection per query and thus no efficiency requirements.}

\begin{figure*}[t]
    \centering
    \includegraphics[trim={0.5cm 0.5cm 0cm 0cm},width=\linewidth]{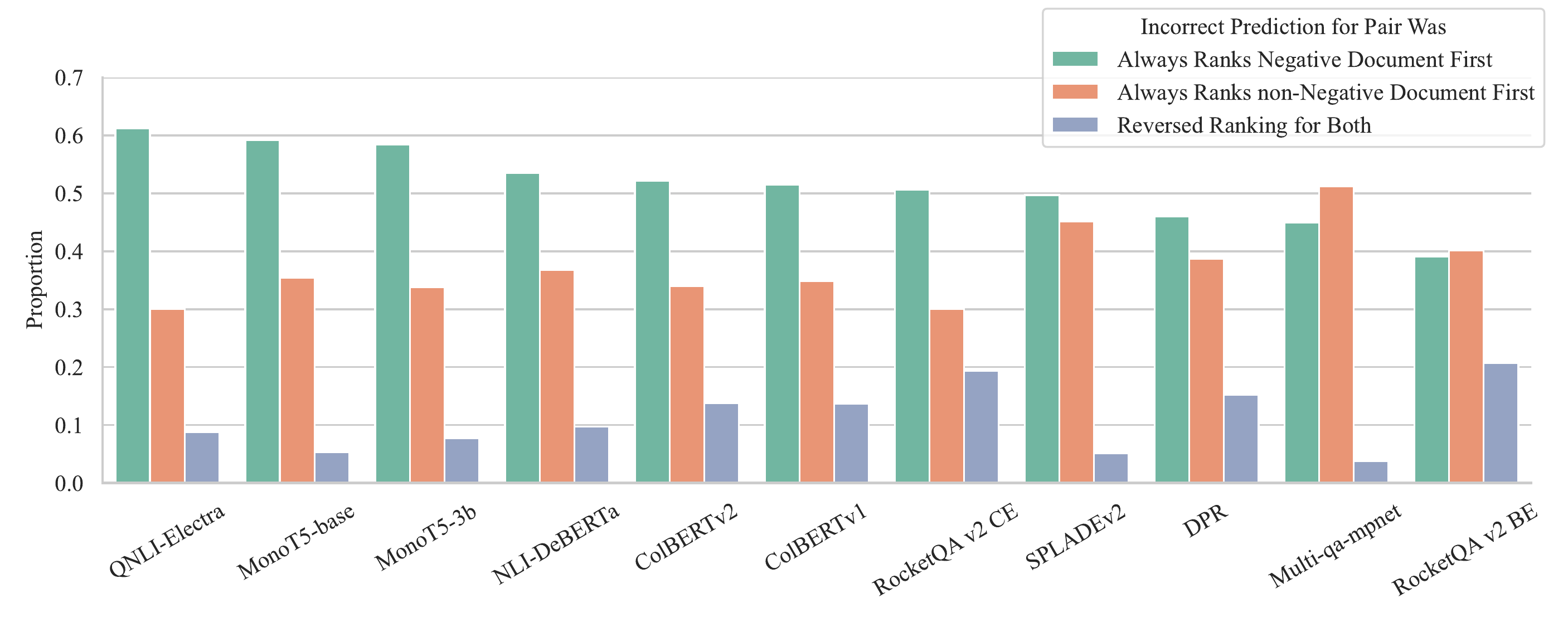}
    \caption{Error analysis of the model predictions, detailing whether models preferred (e.g. by ranking first for both queries) the document with negation (green), the edited non-negation document (orange), or predicted the reversed ranking for both queries (blue). Models that performed better generally preferred negation documents when they made incorrect predictions while bi-encoder models were more balanced in their errors.\label{fig:preference}\vspace{-1em}}
\end{figure*}

\paragraph{Bi-Encoders} Another common category of IR models are bi-encoders, which embed both documents and queries into a single vector representation. At inference time the similarity is computed via a simple dot product or cosine similarity. Due to the popularity of this category, we include a broad spectrum: models from SentenceTransformer \cite{reimers-2019-sentence-bert} trained on MSMarco and/or Natural Questions, DPR \cite{karpukhin-etal-2020-dense}, CoCondenser \cite{gao2021unsupervised}, and RocketQA \cite{qu2020rocketqa,ren2021rocketqav2}. Note that these models span a wide variety of pre-training tasks, base models, and complex training/additional fine-tuning strategies like hard negative mining and distillation.

\paragraph{Cross-Encoders} Cross-encoders encode both the document and query at the same time, computing attention across both pieces of text. This type of representation is the most expressive but also the most time-intensive, especially for larger models. We use various SentenceTransformer cross-encoders including those trained on MSMarco and various NLI datasets \cite{demszky2018transforming,N18-1101,cer2017semeval}, RocketQAv2 cross-encoders \cite{qu2020rocketqa,ren2021rocketqav2}, as well as MonoT5 cross-encoders \cite{nogueira-etal-2020-document}. Note that MonoT5 models are significantly larger (up to 33x larger for 3B) and more expensive than the other cross-encoders.\footnote{T5 models are also typically used for generative retrieval (GR) \cite{tay2022transformer}; thus we do not evaluate GR methods since (1) T5 is evaluated with MonoT5 already and (2) GR has been shown to be unable to scale to standard-sized collections \cite{pradeep2023does} and is not used in practice.}

\paragraph{Random}
We include a baseline that randomly ranks the two documents. Since there are two pairs, the expected mean pairwise accuracy is 25\% ($\frac{1}{2} * \frac{1}{2}$).

\section{Results}
\subsection{Main Results}
The main results are presented in Table~\ref{tab:main}. We see that the more expressive the representation, the better the models generally perform. 

No bi-encoder architecture scores higher than 12\% paired accuracy despite the method of pre-training (e.g. CoCondenser) or the type of contrastive training data (MSMarco, NQ, etc.) with most models performing in the 5-10\% range.

In the sparse category, we see that TF-IDF scored only 2\% paired accuracy. Since we did not allow annotators to use words that were in only one of the paragraphs, this is to be expected.\footnote{Note that the 2\% performance, instead of 0\%, is due to our annotation interface not restricting partial matches (e.g. `version" vs ``versions", ``part" vs ``parting" etc.).} For neural sparse models, all SPLADEv2++ models perform similarly to the bi-encoders, at around 8\% paired accuracy.

The late interaction style models perform significantly better than bi-encoders and sparse models, with ColBERTv1 scoring 19.7\% and ColBERTv2 scoring 13.0\%. Due to the nature of this model we are able to visualize the MaxSim operator to understand its performance (Section~\ref{sec:colbert}).

The cross-encoder models performed the best, with MonoT5 (the default ``base" version) performing at 34.9\% paired accuracy (and the largest version at 50.6\%). Interestingly, the cross-encoders trained on NLI datasets generally performed better than cross-encoders trained on MSMarco, likely due to the fact that MSMarco contains little negation while NLI datasets typically do have negation.

Overall, despite the strong scores of these models on various standard IR benchmarks, nearly all models perform worse than randomly ranking.  Only a handful of cross-encoder models perform better, and they are the slowest and most expensive category of retrieval models. Even these models however, perform significantly below humans and have far from ideal performance.

\subsection{How does model size affect the results?}
We note that Table~\ref{tab:main} includes different sizes of MonoT5. We see that as model size increases, so does the accuracy (from around 28\% with MonoT5-small to around 51\% for MonoT5-3B). This aligns with results shown in the natural language processing community about model size \cite{mckenzie2022round1,wei2022inverse,ravichander2022condaqa,weller2023according}.

However, unlike NLP, IR is typically more latency constrained. Thus, models like MonoT5-3B are only feasible for re-ranking and not for first-stage retrieval (c.f.  Section~\ref{sec:discussion} for more discussion).

\subsection{ColBERT analysis}   
\label{sec:colbert}
As ColBERT models provide token-level vectors and use the MaxSim operator, we are able to visualize whether the max operator pays attention to the negation words (Figures~\ref{fig:colbert} and \ref{fig:colbert_with_negation} in the appendix, due to space constraints). We find in all sampled instances that the MaxSim operator in ColBERTv1 ignores negation words, not selecting them as the max for any query token. Thus, with default training this is a crucial flaw when it comes to processing negation, which causes its less-than-random performance. However, it is possible to fine-tune these representations to put more weight on the negation words so that the MaxSim correctly identifies them, as seen in Section~\ref{sec:training}.

\subsection{Error Analysis}
\label{sec:preference}
We conduct an error analysis to determine which document models prefer for a given pair. Models can prefer (e.g. rank highest in both queries) the document with negation, the edited non-negation document, or predict the reversed rank for both queries. Figure~\ref{fig:preference} shows that the models trained on NLI (and cross-encoders) greatly preferred the document with negation, while bi-encoder models tended to prefer them equally.  Reversed rankings are uncommon, with bi-encoder models having the highest percentage (e.g. RocketQA at $\sim$20\%).

\begin{figure*}
\centering
\begin{subfigure}[b]{1\textwidth}
       \includegraphics[trim={0cm 0.5cm 0cm 0cm},width=\linewidth]{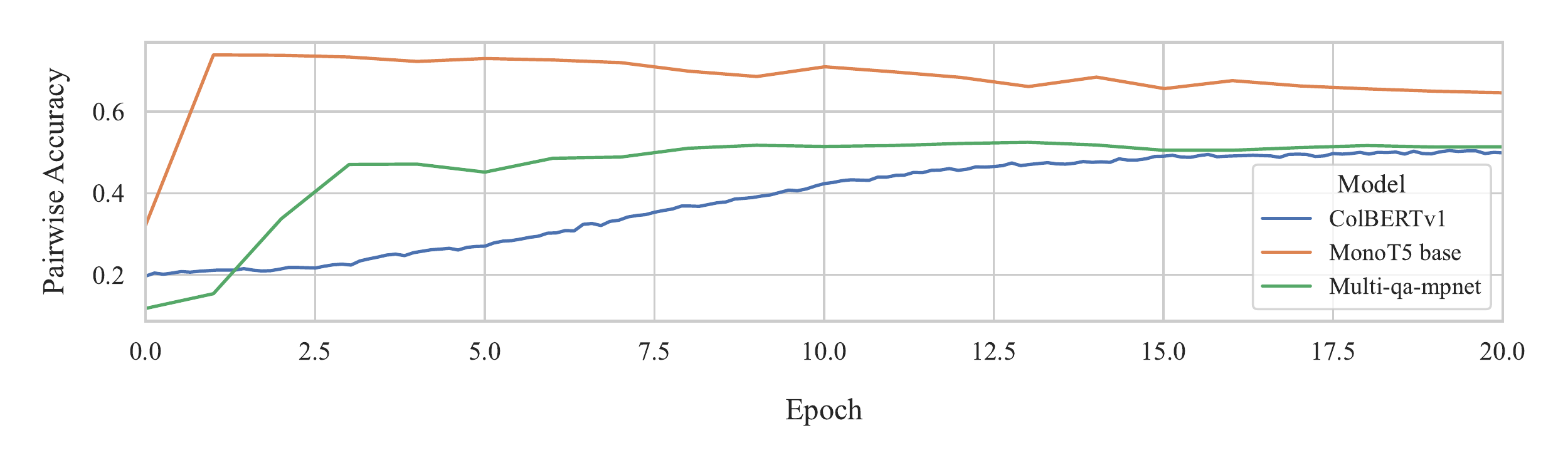}
       \label{fig:training_plot}
\end{subfigure}

\begin{subfigure}[b]{1\textwidth}
    \includegraphics[trim={0cm 1.5cm 0cm 2cm},width=\linewidth]{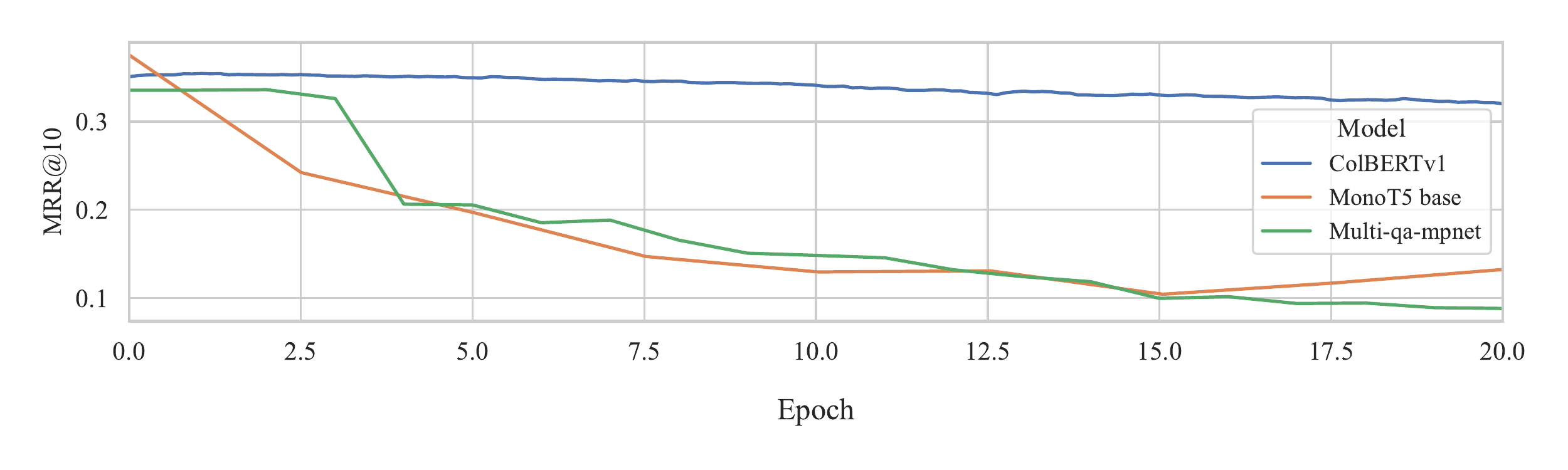}
   \label{fig:training_plot_msmarco}
   \end{subfigure}

\caption{How fine-tuning on \datasetname{}'s training set affects results on NevIR and MSMarco: upper shows \datasetname{}'s pairwise accuracy scores on test while training for up to 20 epochs, lower shows MSMarco dev MRR@10 scores. For QNLI-electra-base see Appendix~\ref{app:qnli}.\label{fig:training_plot_all}\vspace{-1.25em}}
\end{figure*}

\section{Fine-Tuning on \datasetname{}}
\label{sec:training}
Table~\ref{tab:main} shows that models trained on standard IR training datasets do not show strong results on \datasetname{}. However, none of the standard IR datasets include much negation in their queries (potentially due to production systems biasing users, c.f. Section~\ref{sec:discussion}). Thus, in this section we fine-tune IR models on \datasetname{}'s training set to see how negation-specific training data improves performance. 

We use the top performing model from non-sparse categories: multi-qa-mpnet-base-dot-v1 from SentenceTransformers, ColBERTv1, and MonoT5-base from PyGaggle. We fine-tune them using SentenceTransformers, the original ColBERTv1 code, and the original PyGaggle code. We train for 20 epochs and evaluate them on \datasetname{} test and MSMarco dev after each epoch.

Figure~\ref{fig:training_plot_all} shows that fine-tuning on negation data improves performance significantly, but still leaves a large gap to perfect (and the human score of) 100\% paired accuracy. As would be expected, the large MonoT5 model quickly learns and then overfits to the data (while quickly losing performance on MSMarco). Interestingly, ColBERT takes much longer to learn (due to the MaxSim operator), slowly increasing over nearly 20 epochs to learn what the bi-encoder model quickly learned in less than 3. However, we find that ColBERT has a much lower and slower drop in ranking scores on MSMarco (Figure~\ref{fig:training_plot_all} lower). We show visualizations of the MaxSim operator before and after \datasetname{} training in Appendix~\ref{app:colbert}, illustrating that before training the MaxSim operator ignores negation, while after training it learns to correctly include it.

\section{Discussion and Implications}
\paragraph{Implication for Current Systems}\label{sec:discussion}
IR model's performance on \datasetname{} indicates that first stage retrievers do not take negation into account when doing retrieval. Thus, to perform well on negation with current models, expensive cross-encoder re-rankers are necessary but not sufficient to achieve good results. Furthermore, our analysis indicates that in order to best learn negation (and significantly improve their performance), models should incorporate negation into their training data.

Thus, when high precision for negation retrieval is \textit{not needed} (e.g. some first stage retrieval settings), current models may be effective, as they will retrieve lexically similar documents regardless of negation. However, in order to have \textit{high-precision} retrieval with negation (and documents with both negation and non-negation have high lexical overlap), expensive cross-encoders are the only current models that perform better than random ranking. \datasetname{} provides the only dataset for measuring and improving retrieval with negation.

\paragraph{Implications for Current Users} 
Anecdotally, most users tend to avoid using negation queries in production IR systems like Google Search. This may be a self-reinforcing problem, as users have found poor results when they use negation in search and hence avoid using negations in the future. For example, the webpage for the University of Utah article that is shown in Figure~\ref{fig:tweet} has since been updated and currently includes no negation words.

Thus, it is unclear whether queries with negation are less common because of people's actual information needs or because production systems have biased users (and content creators) into an avoidance of negation. We hope that by introducing a benchmark for IR evaluation we can help enable these types of queries in the future.

\section{Conclusion}
We proposed to benchmark negation in neural information retrieval and built a benchmark called \datasetname{} to explore this problem, crowdsourcing annotations from Mechanical Turk.
We found that modern IR models perform poorly on this task, with cross-encoder models performing the best (slightly above random performance) and all other architectures (bi-encoder, sparse, and late-interaction) performing worse than random.
Further we showed that simply including negation in fine-tuning provides significant gains, although there is still room for improvement to reach human performance.
We hope that this benchmark inspires future work into improving information retrieval model's ability to recognize negation.

\section{Limitations}
Our work provides results for a broad range of IR models (including the most common and popular), but does not provide results for all possible IR models due to space and time. We welcome future research into investigating alternative methods and models to improve performance on \datasetname{}.

Our dataset follows previous work in designing contrastive evaluation datasets \cite{kaushik2019learning,penha2022evaluating,macavaney2022abnirml} and we note that because of this our work does not provide a large-scale collection to go along with our queries (enabling an analysis of recall along with the precision we measure), as might be found in classic IR datasets. However, as shown by a large body of work (see Section~\ref{sec:contrastive}), contrastive evaluations can provide important insight into understanding and improving neural models. We leave large collection creation with negation and analysis of recall performance to future work.

\section*{Acknowledgements}
OW is supported by the National Science Foundation Graduate Research Fellowship Program.

\bibliography{anthology,custom}

\appendix

\section{Annotation Interface}
\label{app:annotation_interface}
\begin{figure*}[t]
    \centering
    \includegraphics[trim={0cm 0.25cm 0 0.25cm},width=\linewidth]{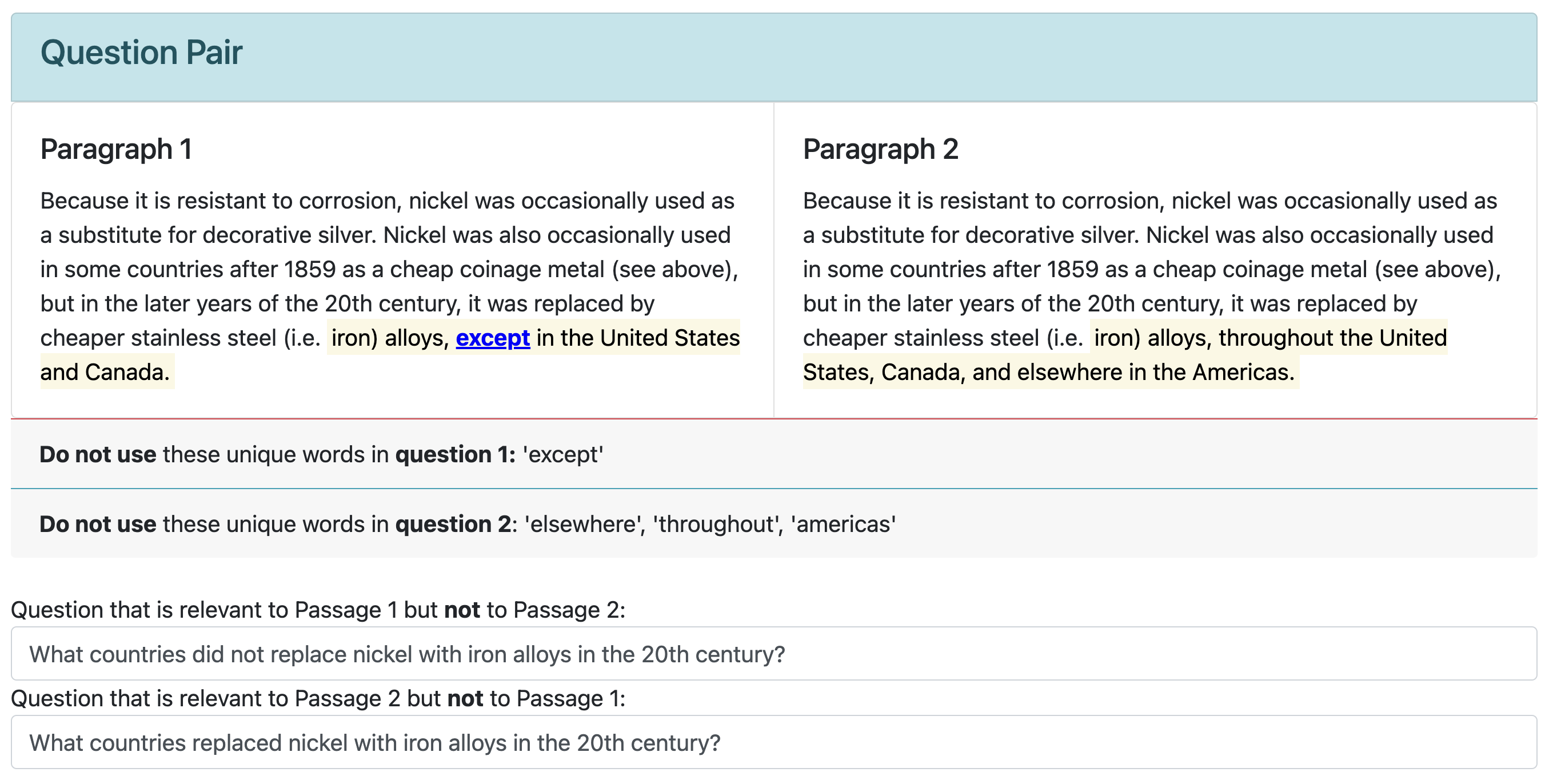}
    \caption{Number of unique words between the two queries.\vspace{-0.25em}\label{fig:annotation_interface}}
\end{figure*}

In Figure~\ref{fig:annotation_interface} we show the annotation interface provided to workers on Mechanical Turk.

\section{Document Edit Types}
\label{app:edit_types}
\begin{figure}[t]
    \centering
    \includegraphics[trim={0.25cm 0cm 0cm 0cm},width=\linewidth]{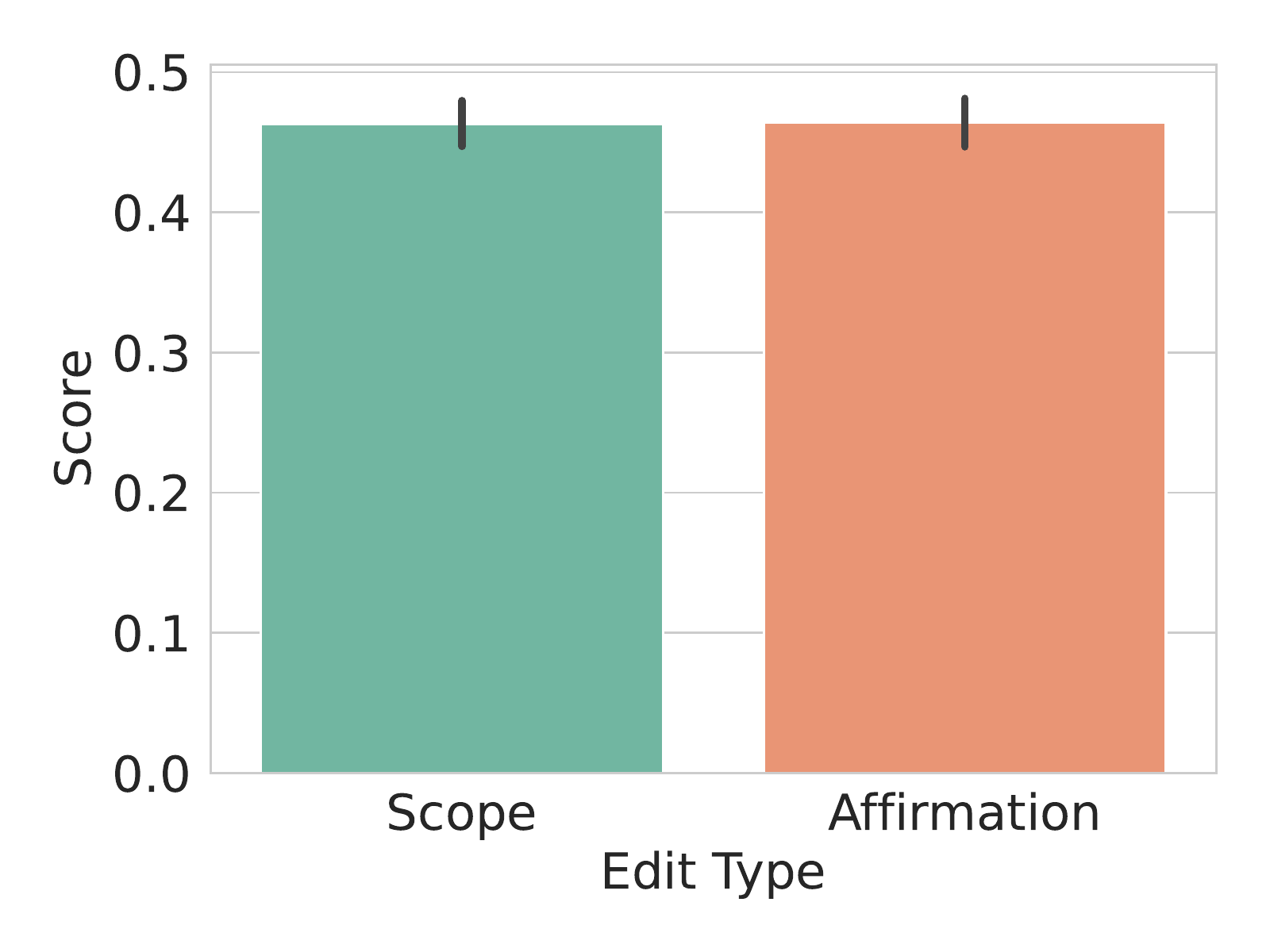}
    \caption{Edit types from the CondaQA dataset and their average pairwise scores. Error bars indicate a 95\% confidence interval. \label{fig:edit_type}}
\end{figure}

We also analyze the edit types from the original CondaQA dataset to see if they impact the pairwise accuracy. We see in Figure~\ref{fig:edit_type} that there is no statistical difference (given the 95\% confidence interval) between the two types of edits for the MonoT5-3B model (and we note that other models are similar and hence we only include one model).

\section{Cosine Similarity after Fine-Tuning}
\label{app:dense_train}
\begin{figure*}[t]
    \centering
    \includegraphics[trim={0cm 0.75cm 0cm 0cm},width=\linewidth]{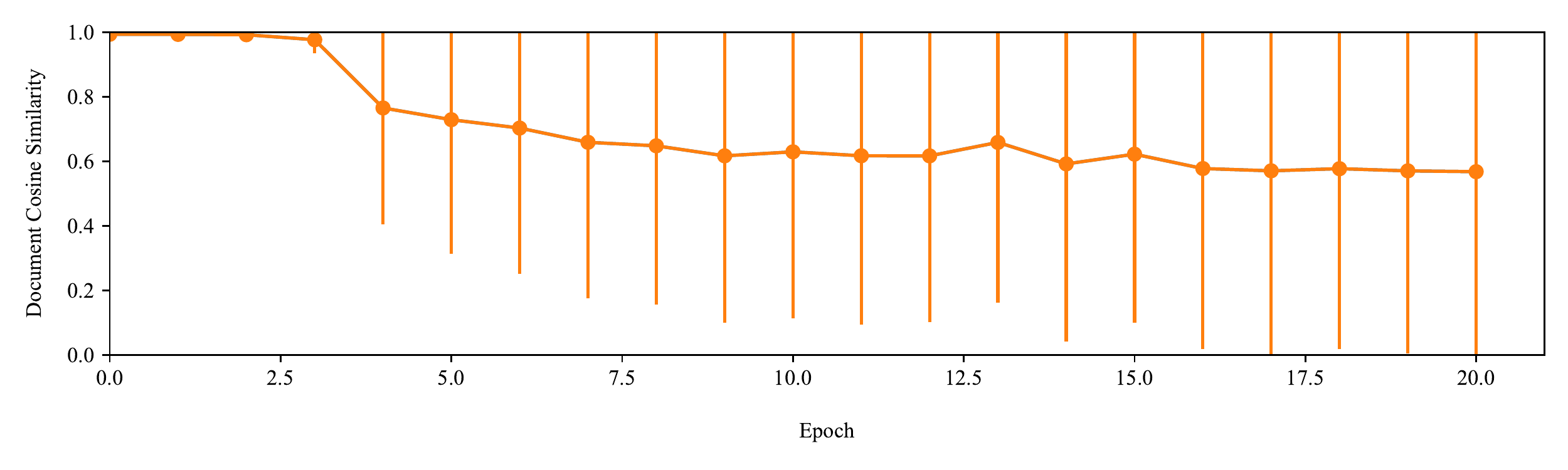}
    \caption{Cosine similarity scores between documents in the pairs during fine-tuning for the multi-qa-mpnet bi-encoder model. Error bars indicate one standard deviation.\label{fig:sim_score}}
\end{figure*}

In Figure~\ref{fig:sim_score} we see the results for cosine similarity between each document pair during different epochs. We can see that the representations start nearly identically, but shift to be further apart and to have more variance as training continues. This plot was created using the multi-qa-mpnet model, but other dense models show similar results.

\section{ColBERT Analysis}
\begin{figure*}[t]
    \centering
    \includegraphics[trim={0cm 0cm 2cm 0cm},width=\linewidth]{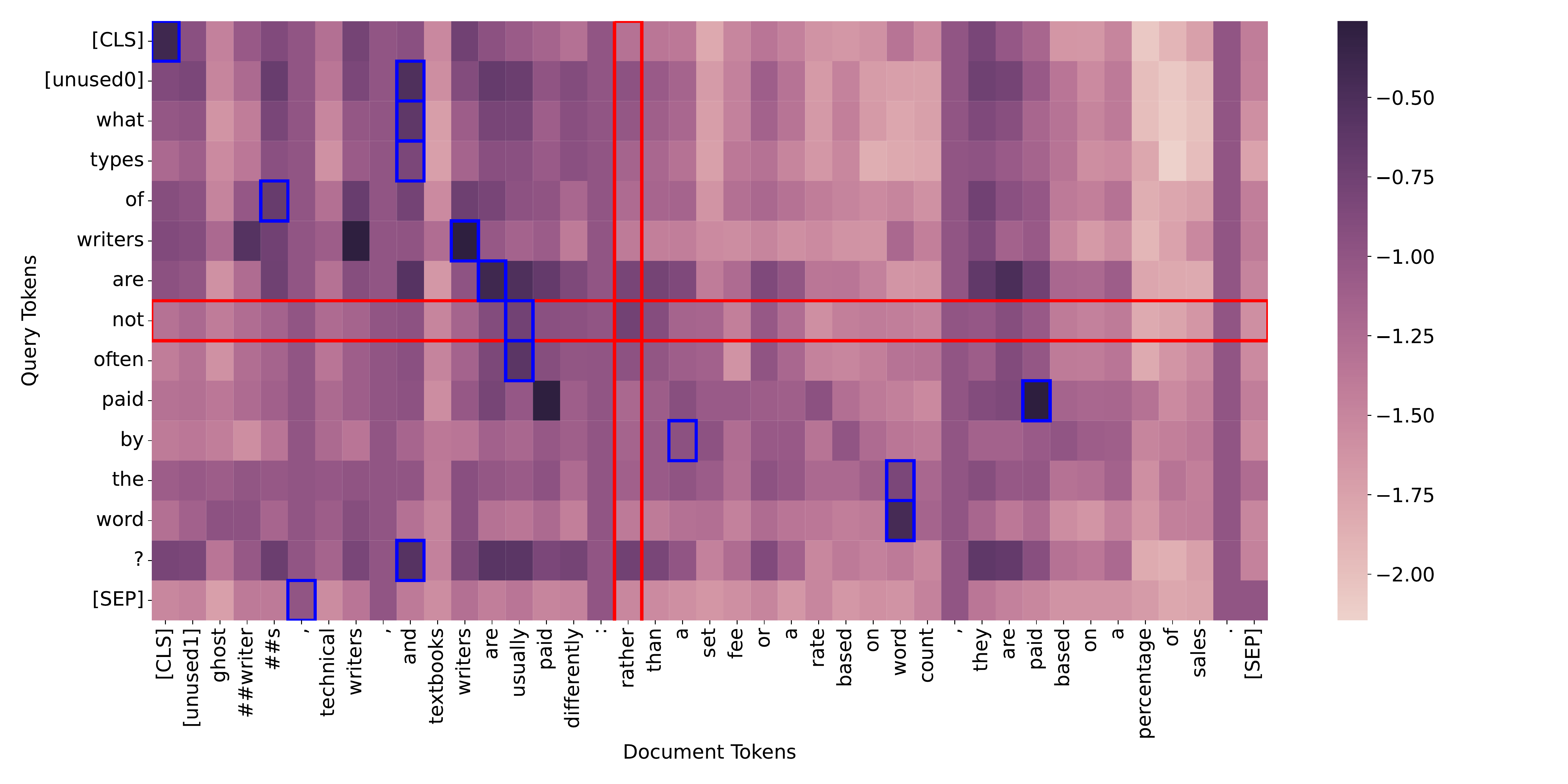}
    \caption{An example instance with results from ColBERT's MaxSim operator from the ColBERTv1 model. Red highlights indicate the tokens corresponding to the negation (or lack of negation) while blue highlights indicate the max token for the MaxSim operator. Note that this model predicts the MaxSim token of ``usually" for ``not" and has no Max for the crucial word ``rather". However, further fine-tuning helps improve this, see Figure~\ref{fig:colbert_with_negation}.\label{fig:colbert}}
\end{figure*}

We show two heatmaps for ColBERTv1 models, the first using the original model trained on MSMarco and the 2nd after fine-tuning for 20 epochs on \datasetname{}. We see in Figure~\ref{fig:colbert} that the model fails to associate any maximum tokens with the crucial word ``rather" instead associating ``not" with ``usually". In contrast, after training on \datasetname{}, the model correctly associates ``rather" with ``not".

\label{app:colbert}
\begin{figure*}[t]
    \centering
    \includegraphics[trim={0cm 0cm 2cm 2cm},width=\linewidth]{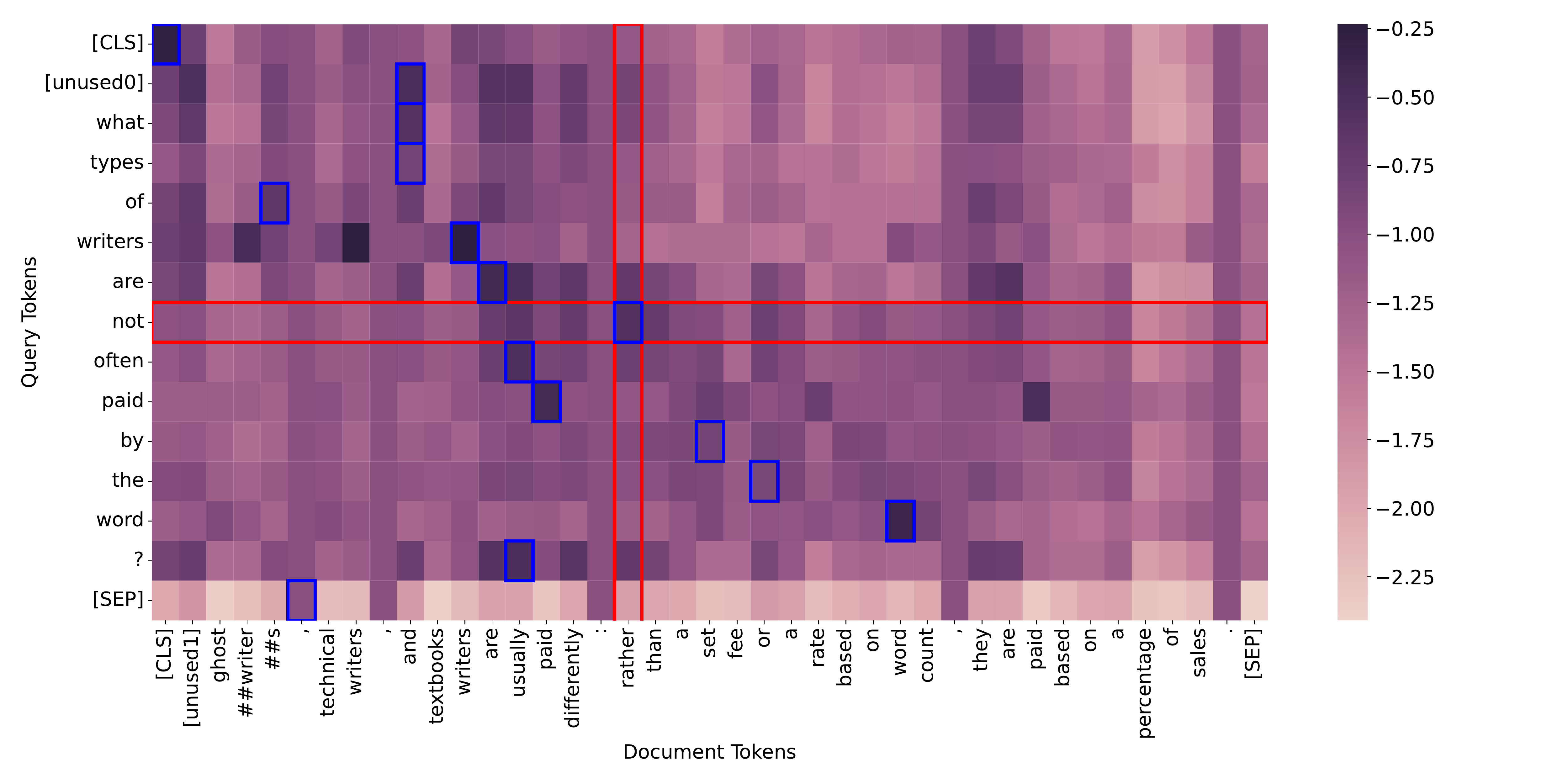}
    \caption{An example instance with results from ColBERT's MaxSim operator from the ColBERTv1 model trained for 20 epochs on \datasetname{}. Red highlights indicate the tokens corresponding to the negation (or lack of negation) while blue highlights indicate the max token for the MaxSim operator. Note that this model correctly associates the word ``not" with the crucial word ``rather" unlike Figure~\ref{fig:colbert}.\label{fig:colbert_with_negation}}
\end{figure*}

\begin{figure*}
\centering
\begin{subfigure}[b]{1\textwidth}
       \includegraphics[trim={0cm 1.5cm 0cm 0cm},width=\linewidth]{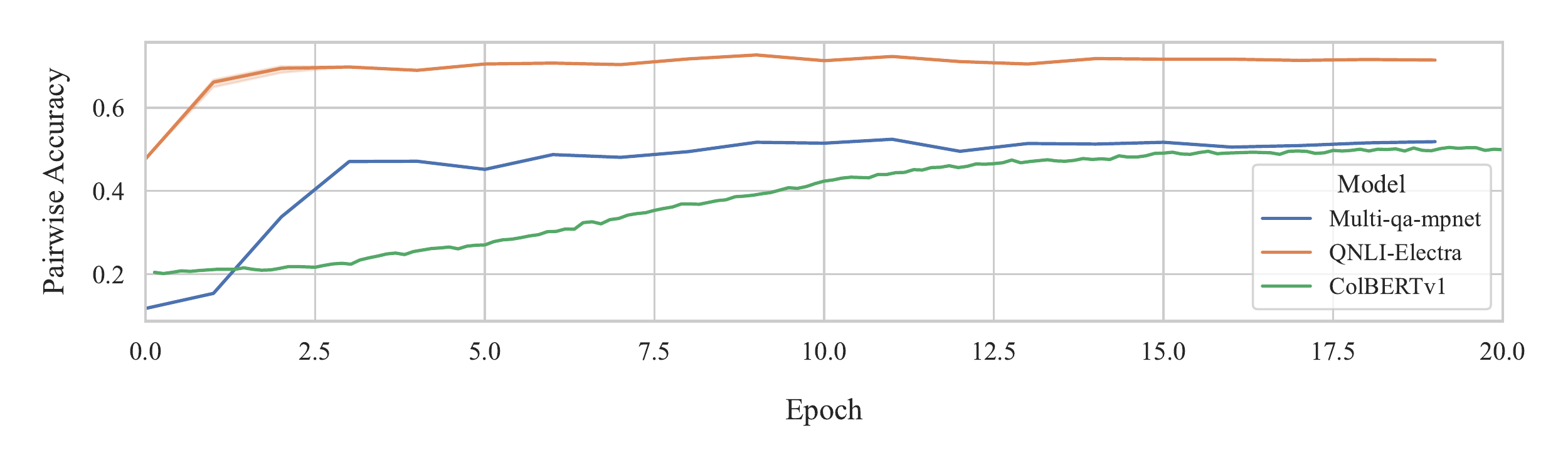}
       \label{fig:training_plot_qnli}
\end{subfigure}

\caption{Results from fine-tuning IR models on the \datasetname{} training set, including QNLI-electra-base. The plot shows \datasetname{} test set pairwise accuracy scores while training for up to 20 epochs\label{fig:training_plot_all_app}}
\end{figure*}

\section{Results with training QNLI-electra-base on \datasetname{}}
\label{app:qnli}
Figure~\ref{fig:training_plot_all_app} shows results with QNLI-electra-base also, which shows similar results to MonoT5 in the main paper. We do not show results for MSMarco as QNLI-electra-base was not trained on MSMarco.

\section{Importance of Negation in Retrieval}
\label{app:twitter}
\begin{figure*}[t]
\centering
    \tcbox[size=fbox]{\includegraphics[trim={0cm 0cm 0cm 0cm},width=0.45\linewidth]{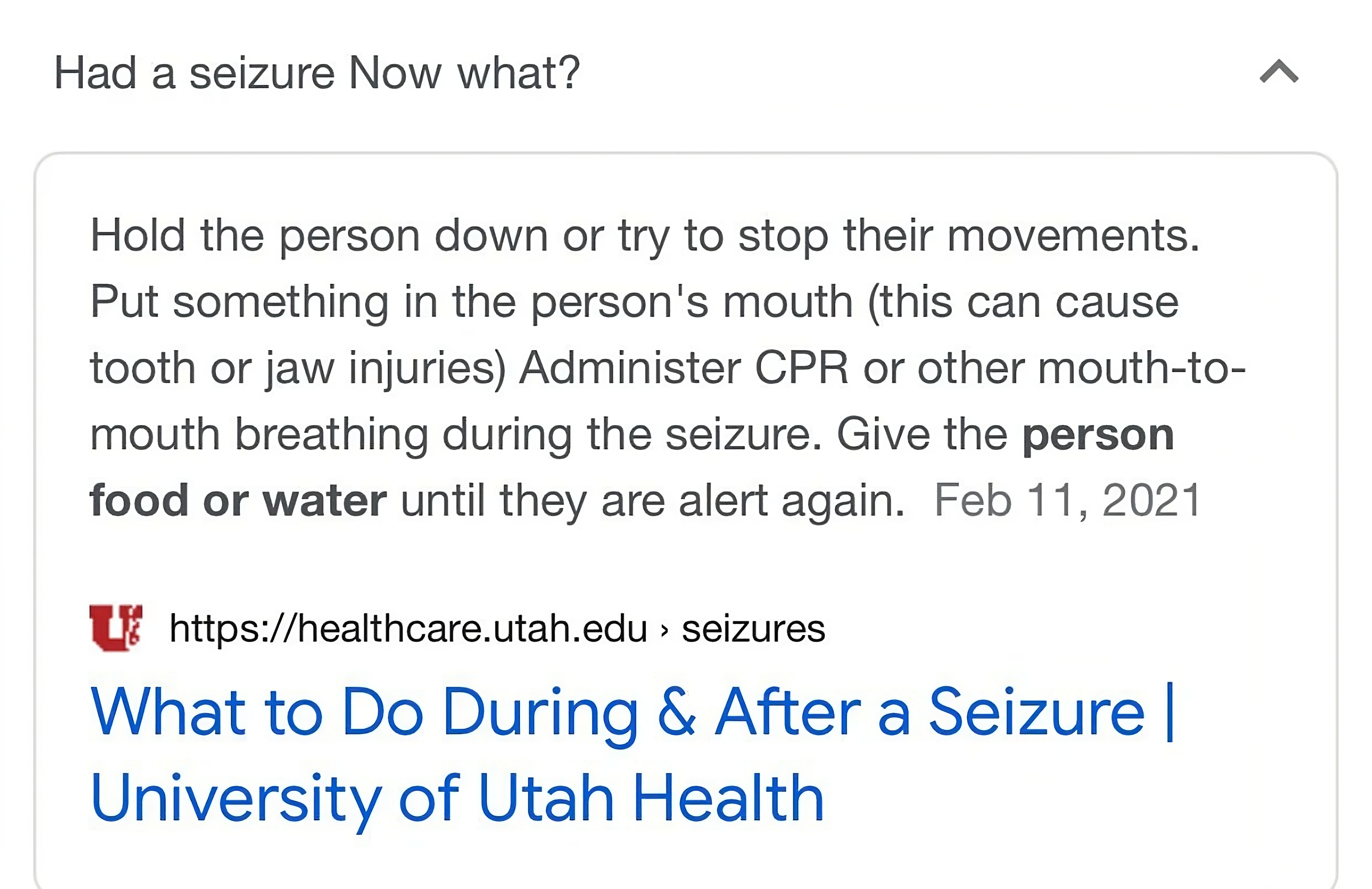} \vspace{20cm}
    \includegraphics[trim={0cm 0cm 0cm 0cm},width=0.45\linewidth]{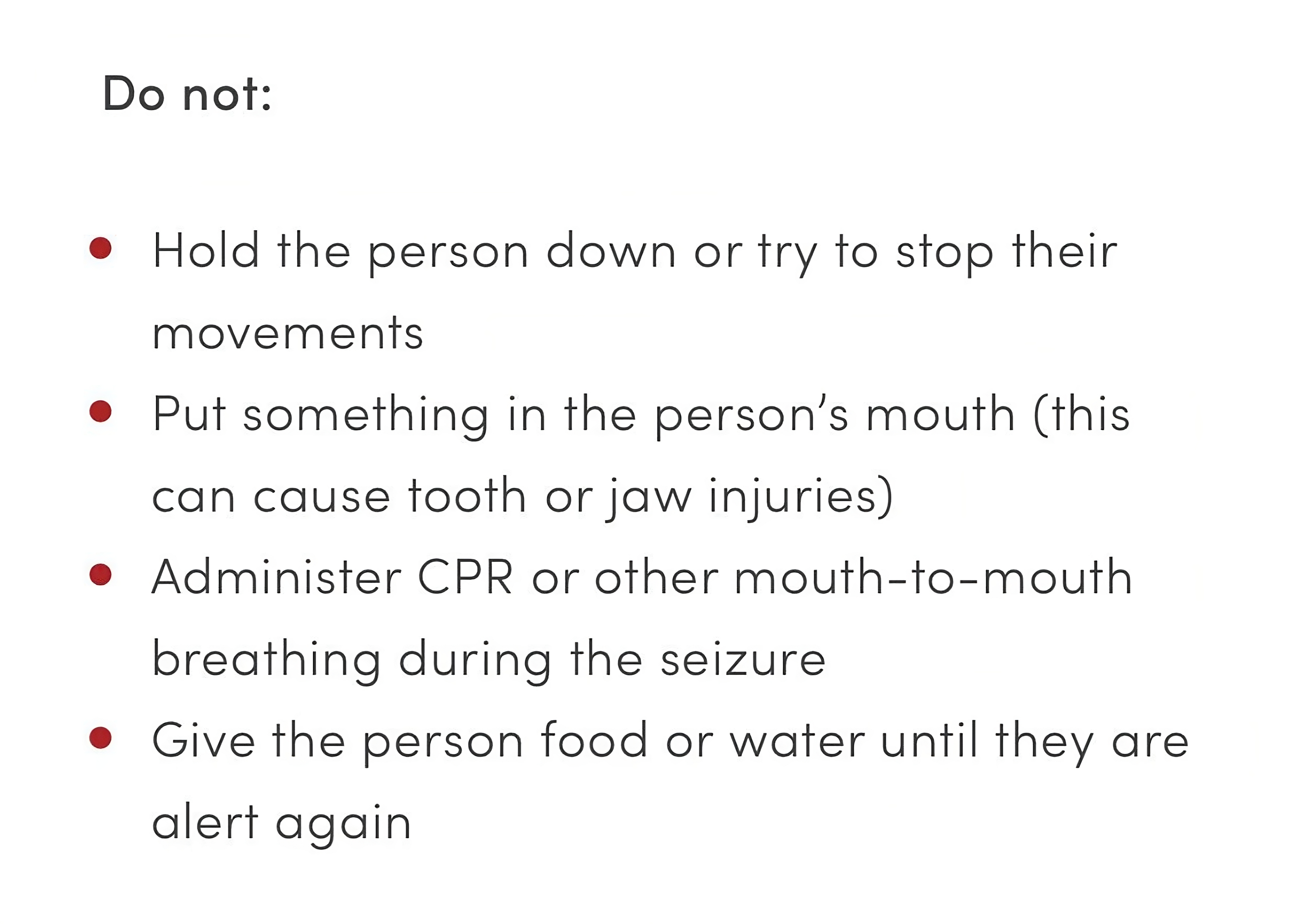}}
    \caption{Reproduction of the tweet showing Google Search making a life-threatening recommendation and failing to catch the negation in the article.\label{fig:tweet_app}}
\end{figure*}

We include pictures of the tweet referenced at \url{https://x.com/soft/status/1449406390976409600} in Figure~\ref{fig:tweet_app}, showing the dangers of not understanding negation.

\section{Hyperparameters and Computational Resources}
All experiments were run on a cluster of V100s with each experiment taking less than an hour on one V100.

We use default hyperparameters for all models for inference (and many models do not have any hyperparameters). For ColBERT training we use their code that has a default learning rate of 3e-6 and for bi-encoder training we use Sentence-Transformers that has a default of 2e-5.

\end{document}